\documentclass[pra,twocolumn,numerical]{revtex4-1}

\newcommand{\ket}[1]{| #1 \rangle}
\newcommand{\bra}[1]{\langle #1 |}

\usepackage{amsmath,amssymb}
\usepackage{acronym}
\usepackage{setspace}

\usepackage{graphicx}

\usepackage{overpic}

\usepackage{subfigure}

\begin{document}


\title{Micro-resonator based all-optical transistor}

\author{B. D. Clader}\email{dave.clader@jhuapl.edu}
\author{S. M. Hendrickson}\affiliation{Research and Exploratory Development Department \\ The Johns Hopkins University Applied Physics Laboratory, Laurel, MD 20723, USA}

\begin{abstract}
We present theoretical estimates for a high-speed, low-loss, all-optical transistor using a micro-resonator device, whose fields interact evanescently with Rubidium vapor.  We use a four-level electromagnetically induced absorption scheme to couple the light fields of the transistor.  We show results indicating that a weak control beam can switch a much stronger signal beam, with contrast of greater than 25 dB and loss less than 0.5 dB.  The switching timescale is on the order of 100 ps.
\end{abstract}


\maketitle

\acrodef{SiN}{Silicon Nitride}
\acrodef{Cs}{Cesium}
\acrodef{Rb}{Rubidium}
\acrodef{TPA}{two-photon absorption}
\acrodef{CTPA}{coherent two-photon absorption}
\acrodef{SPA}{single photon absorption}
\acrodef{SEM}{scanning electron microscope}
\acrodef{cavRes}[$\lambda_{c}$]{cavity resonance}
\acrodef{780Res}[$\lambda_{Rb}^{780}$]{780 atomic resonance}
\acrodef{1529Res}[$\lambda_{Rb}^{1529}$]{1529 atomic resonance}
\acrodef{HF}{Hydrofluoric acid}
\acrodef{Q}{quality factor}
\acrodef{QZE}{quantum Zeno effect}
\acrodef{AZE}{anti Zeno effect}
\acrodef{FOM}{figure of merit}
\acrodef{EIT}{electromagnetically induced transparency}
\acrodef{AOM}{acousto-optic modulator}
\acrodef{EOM}{electro-optic modulator}
\acrodef{SERF}{spin-exchange-relaxation-free}
\acrodef{SQUID}{superconducting quantum interference device}
\acrodef{SERS}{stimulated electronic Raman scattering}
\acrodef{PMT}{photo-multiplier tube}

%
%
%
%
All-optical logic devices show promising potential as a way to lower power dissipation, while increasing switching speeds \cite{1250885}.  In addition, the low losses made possible by these devices may be useful for optical quantum computing devices, where single photon loss mitigation is critical \cite{miller2010optical, Dawes29042005, hu2008picosecond, albert2011cavity}.  Recent advancements in high-Q photonic micro-resonators have pushed the technology forward to the point where chip-scale photonic based logical gates are becoming possible \cite{barclay:131108, ShahHosseini:10, PhysRevA.82.031804}.  

Recently, an all-optical switch design was proposed and demonstrated that showed how one could evanescently couple a micro-resonator to Rb vapor, to create a low-loss all-optical Zeno switch \cite{jacobs2009all, ZenoSwitch}.  One of the fundamental issues limiting this technology is the strength of the optical nonlinearities that couple the signal and control fields.  Electromagnetically induced transparency (\acsu{EIT}) \cite{harris:36} has shown promise as a resource for all-optical devices due to its large nonlinearity, which is enhanced by coherent effects \cite{PhysRevLett.102.203902, Zhang:07}. We have theoretically demonstrated how one could use \ac{EIT} together with a micro-resonator to create an all-optical Zeno switch \cite{clader2012}, where a strong control beam switched a weak signal beam.  The scheme we present here does not require that the controlling field be much stronger than the signal field.  This could allow one to use this technique to switch strong fields with weak ones, interchange the control and switched field, and operate near single photon levels.

\begin{figure}[h!]
\begin{center}
\includegraphics[width=5.0cm]{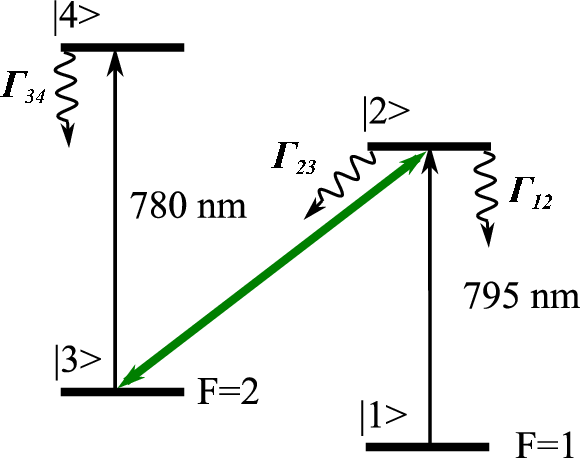}
\caption{\label{fig:NSystem}Four level system, modeled in this paper, including numbering scheme, laser fields, and decay terms.  The two ground states are the $F=1$ and $F=2$ hyperfine levels of the $5^2S_{1/2}$ ground state of $^{87}$Rb.  The excited states are the hyperfine manifolds of the $5^2P_{1/2}$ and $5^2P_{3/2}$ excited states for the $D_1$ (795 nm) and $D_2$ (780 nm) lines respectively. }
\end{center}
\end{figure}

The atomic system we analyze is the four-level scheme proposed in \cite{PhysRevLett.81.3611} and shown schematically in Fig. \ref{fig:NSystem}. It has an \ac{EIT} field that is always on (shown in green) and two signal fields.  If only one signal field is present, the scheme is either in the three-level $\Lambda$ or three-level V \ac{EIT} configuration, such that the \ac{EIT} field renders the medium transparent to the signal field.  However, if both signal fields are present, a \ac{CTPA} mechanism is activated causing both signal photons to be absorbed.  In addition, the various absorption pathways constructively interferere, resulting in a two photon absorption cross section that can be comparable to the single photon absorption cross section.  This effect has been observed experimentally in both cold \cite{Yan:01} and Doppler broadened Rb vapor \cite{Mulchan:00}.  Because of this large nonlinear cross-section, we show that this scheme allows one to create an all-optical transistor in a four-port micro-resonator, that operates below single-photon intensities on average in the cavity.

\begin{figure}
\begin{center}
\subfigure[Single Beam Present]{
\begin{overpic}[width=3.91cm]{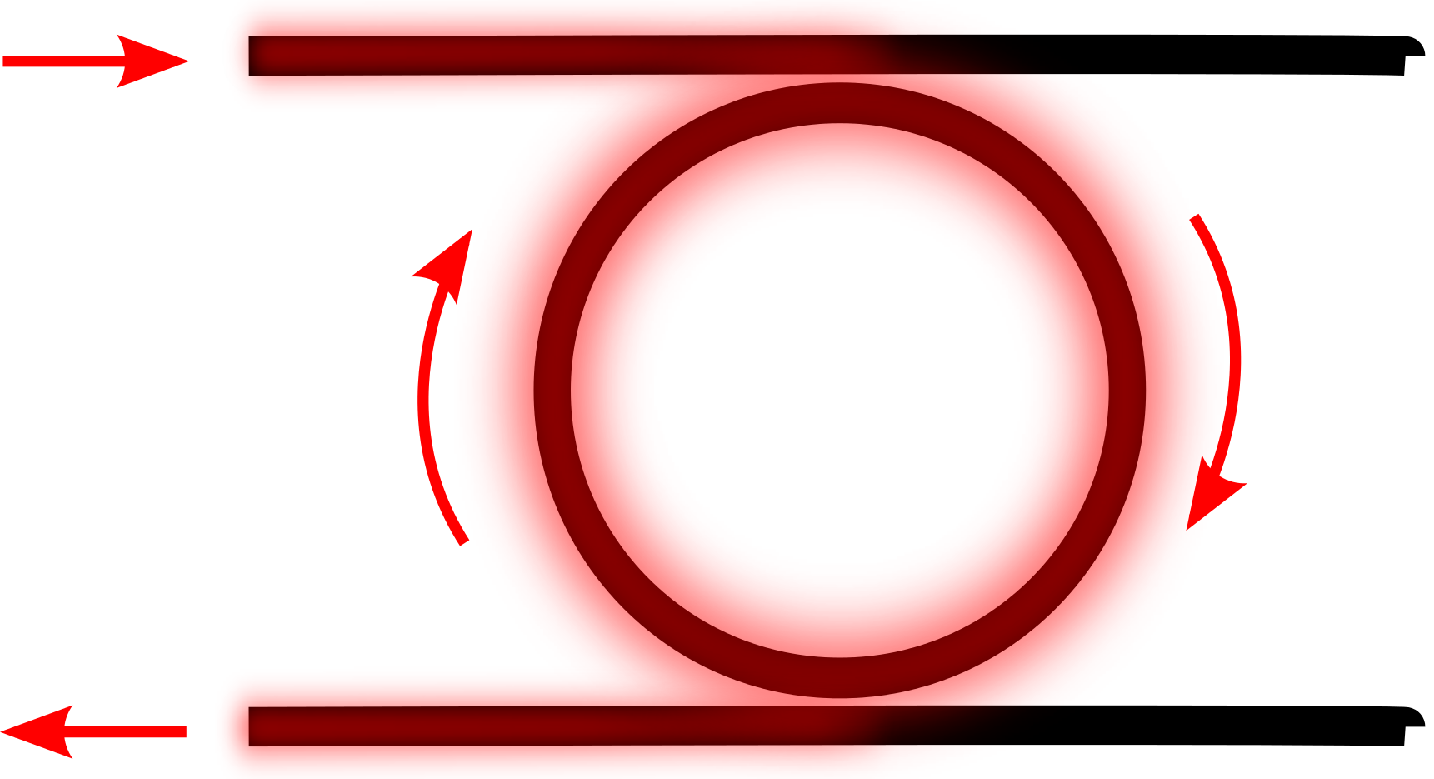}
\put(43,22){EIT}
\put(1,50){$E_{in}$}
\put(1,10){$E_{d}$}
\put(90,50){$E_{t}$}
\end{overpic}
\vspace{2pt}
\label{fig:resonator_spa}
}
%
\subfigure[Two Beams Present]{
\begin{overpic}[width=3.91cm]{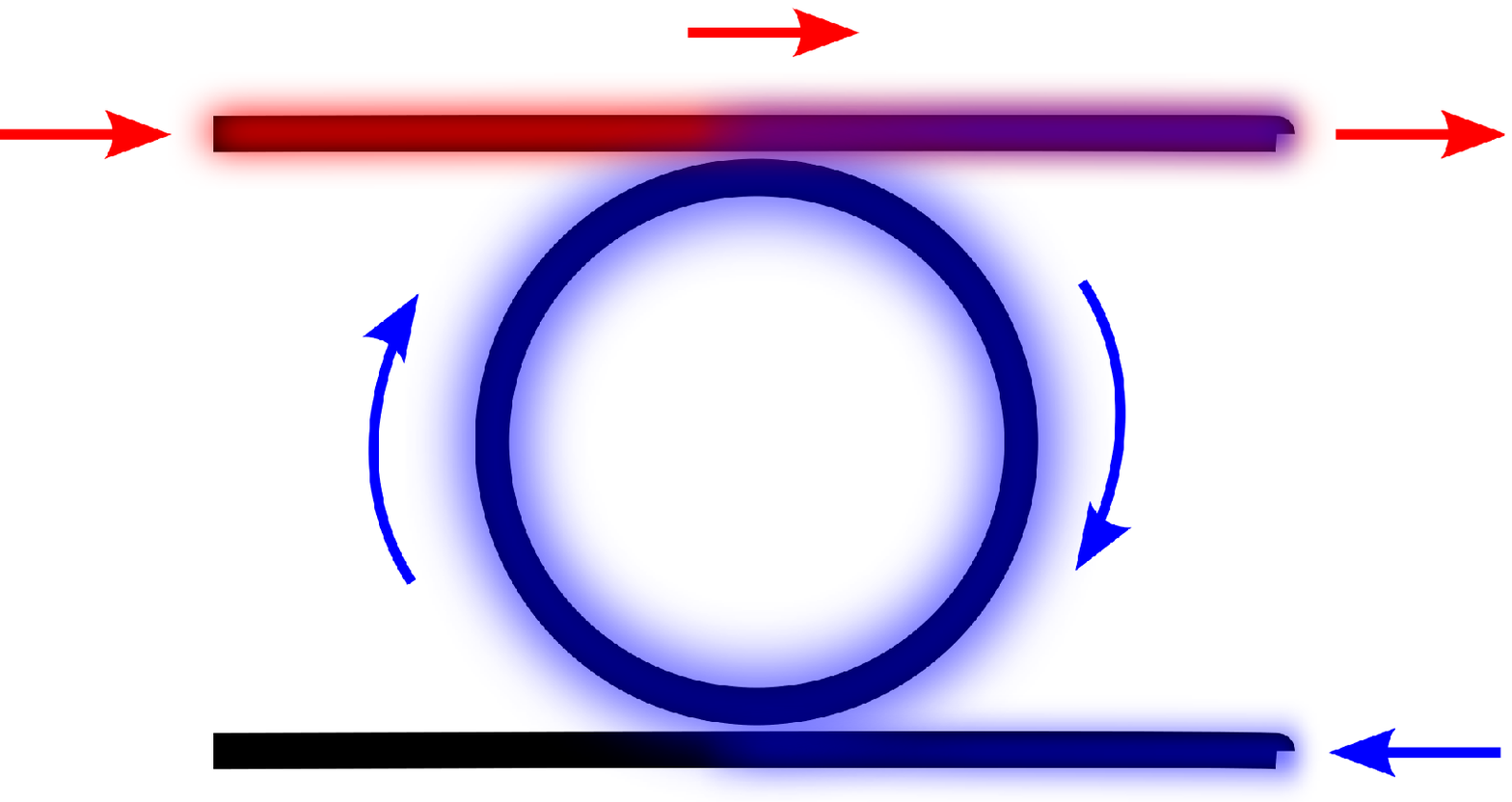}
\put(38,22){CTPA}
\put(1,50){$E_{in}$}
\put(1,10){$E_{d}$}
\put(90,50){$E_{t}$}
\end{overpic}
\label{fig:resonator_eit}
}
\end{center}
\caption{\label{fig:eit_switch}Schematic of the coherent two-photon absorption based all-optical switch.
(a) When only a single input beam, denoted in red, is weakly coupled to the cavity the free-space EIT beam eliminates the evanescent coupling of the signal beam to the atoms surrounding the cavity.  This reduces the external cavity loss, allowing the signal beam to build in the resonator and exit through the drop port $E_d$.
(b) When two beams are present, denoted by blue and red, strong \ac{CTPA} inhibits field buildup in one of the beams causing it to bypass the resonator and exit via the through port $E_t$.  In this way, the presence or absence of the blue beam controls whether or not the red beam couples into the resonator.}
\end{figure}

%
%
%
%

The Hamiltonian for the four-level atomic system in Fig. \ref{fig:NSystem}, interacting with three laser fields is
\begin{align} \nonumber
\label{eq:hamiltonian}
H & =   \Delta_1\ket{2}\bra{2} + (\Delta_1 - \Delta_c)\ket{3}\bra{3} + (\Delta_1 +\Delta_2 - \Delta_c)\ket{4}\bra{4} \\
& - \frac{\Omega_1}{2}\ket{1}\bra{2} - \frac{\Omega_c}{2}\ket{3}\bra{2} - \frac{\Omega_2}{2}\ket{3}\bra{4} + \rm{h.c},
\end{align}
where $\Omega = 2\vec{d} \cdot \vec{\mathcal{E}}/\hbar$ is the slowly varying Rabi frequency, for atomic dipole moment $d$, and slowly varying field amplitude $\mathcal{E}$ of the laser field.  The subscripts 1,2, and c refer to the laser field connecting states $\ket{1} \to \ket{2}$, $\ket{3} \to \ket{4}$, and $\ket{3} \to \ket{2}$ respectively.  The three fields can be detuned from resonance with detuning $\Delta$ subscripted for the appropriate field.

The equations of motion for the atomic density matrix are given by
\begin{equation}
\label{eq:densitymatrix}
i \hbar \frac{\partial \rho}{\partial t} = [H,\rho].
\end{equation}
We add three phenomenological decay terms to Eq. \eqref{eq:densitymatrix} that account for population decay from the excited states into the ground states.  These are denoted in Fig. \ref{fig:NSystem} as $\Gamma_{12}$, $\Gamma_{23}$ and $\Gamma_{34}$.  In addition, we include off-diagonal decoherence decay terms.  We denote these terms as a lower case $\gamma_{ij}$, where the two numerical subscripts refer to the coherence of the state $\ket{i}$ with state $\ket{j}$.  Aside from $\gamma_{13}$ which we take to be very weak, we assume the off-diagonal decoherence rates are dominated by homogeneous terms, and consistent with a positive density matrix \cite{PhysRevA.70.022107, PhysRevA.71.022501}.  They are given by $\gamma_{12} = \frac{1}{2}(\Gamma_{12} + \Gamma_{23})$, $\gamma_{14} = \frac{1}{2}(\Gamma_{12} + \Gamma_{23} + \Gamma_{34})$, $\gamma_{23} = \frac{1}{2}(\Gamma_{12} + \Gamma_{23})$, $\gamma_{24} = \frac{1}{2}(\Gamma_{12} + \Gamma_{23} + \Gamma_{34})$, $\gamma_{34} = \frac{1}{2}\Gamma_{34}$.

As in Ref. \cite{clader2012}, we model the coupling of the waveguides to the cavity using coupled-mode equations \cite{haus1984}.   The through port and drop port transmission rates from coupled-mode theory are
\begin{subequations}
\label{eq:throughdroptransmission}
\begin{align}
T & = \frac{4\Delta^2 + (\kappa_0 + \kappa_e - \kappa_1 + \kappa_2)^2}{4\Delta^2 + (\kappa_0 + \kappa_e + \kappa_1 + \kappa_2)^2} \\
D & = \frac{4\kappa_1 \kappa_2}{4\Delta^2 + (\kappa_0 + \kappa_e + \kappa_1 + \kappa_2)^2},
\end{align}
\end{subequations}
where $\Delta$ is the detuning of the field from the cavity resonance, $\kappa_0$ is the intrinsic cavity linewidth, $\kappa_e$ is related to the absorption rate of the atoms, and $\kappa_1$ and $\kappa_2$ are the cavity -- waveguide coupling rates for waveguides 1 and 2 respectively.  The external loss rate $\kappa_e$ is related to the rate of atomic absorption.  By optically controlling this we can modify the through-port and drop-port transmission rates, resulting in the ability to switch between them.  We assume that both the 780 nm and 795 nm fields are simultaneously resonant in the cavity, by judicious choice of the free-spectral range.

%
%
%
%

\begin{figure}[h!]
\begin{center}
\subfigure[Through Port: $P_1/P_2 = 1$]{
\begin{overpic}[width=3.99cm]{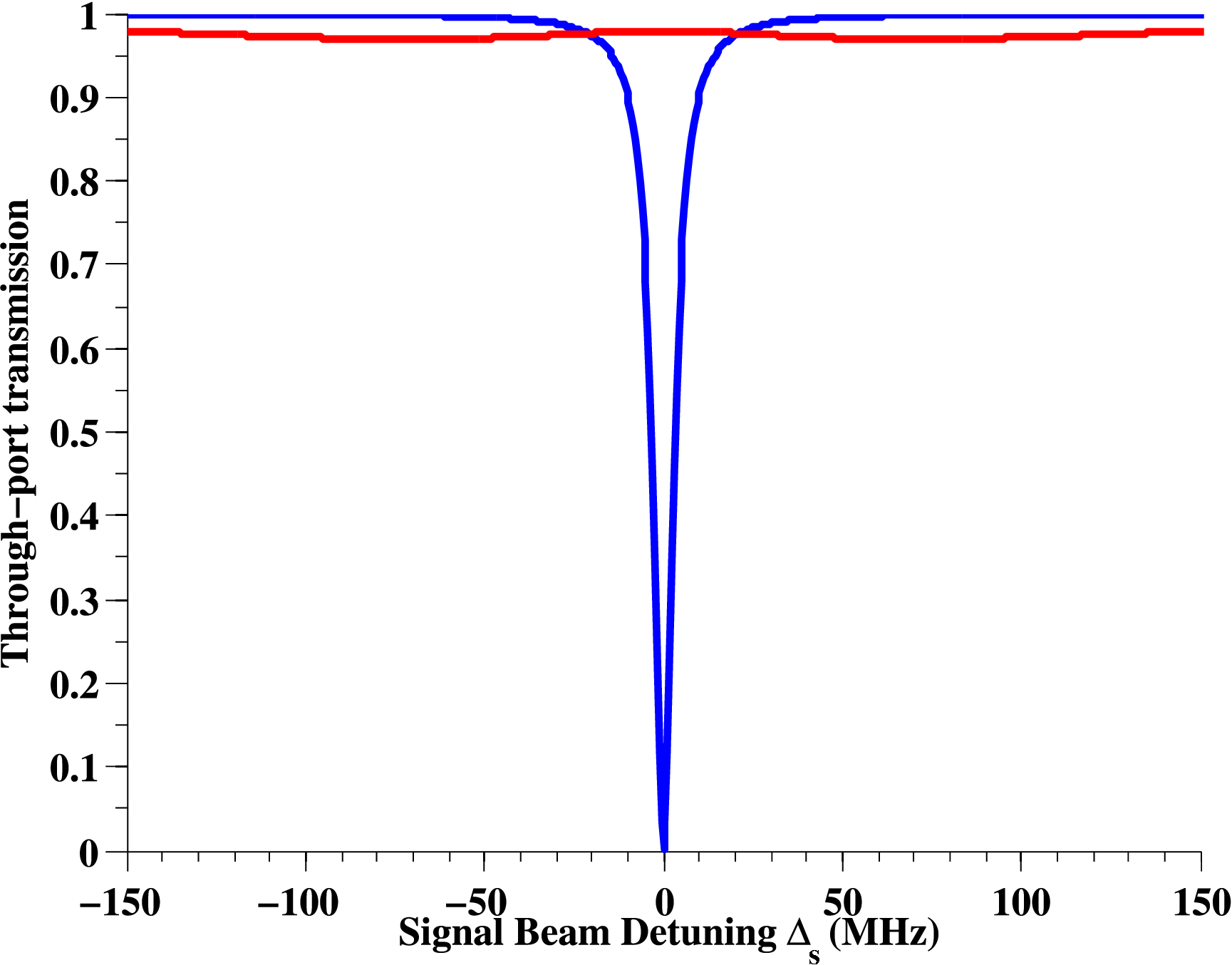}
\put(40,85){{\bf 795 nm Control - 780 nm Signal}}
\end{overpic}
\label{fig:LowerControlEqual-ThroughPort}
}
\subfigure[Drop Port: $P_1/P_2 = 1$]{
\begin{overpic}[width=3.99cm]{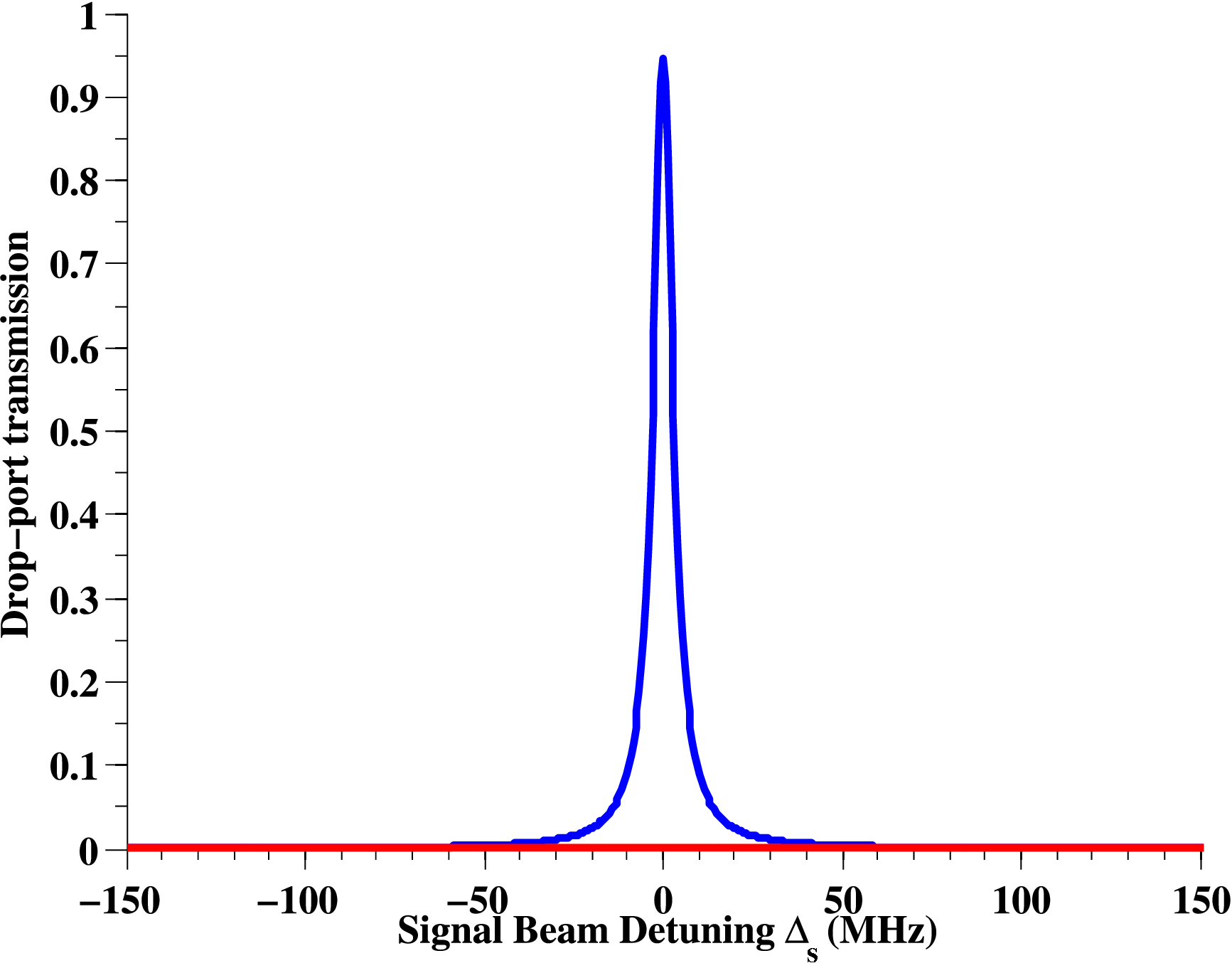}
\end{overpic}
\label{fig:LowerControlEqual-DropPort}
}
\subfigure[Through Port: $P_1/P_2 = 0.1$]{
\begin{overpic}[width=3.99cm]{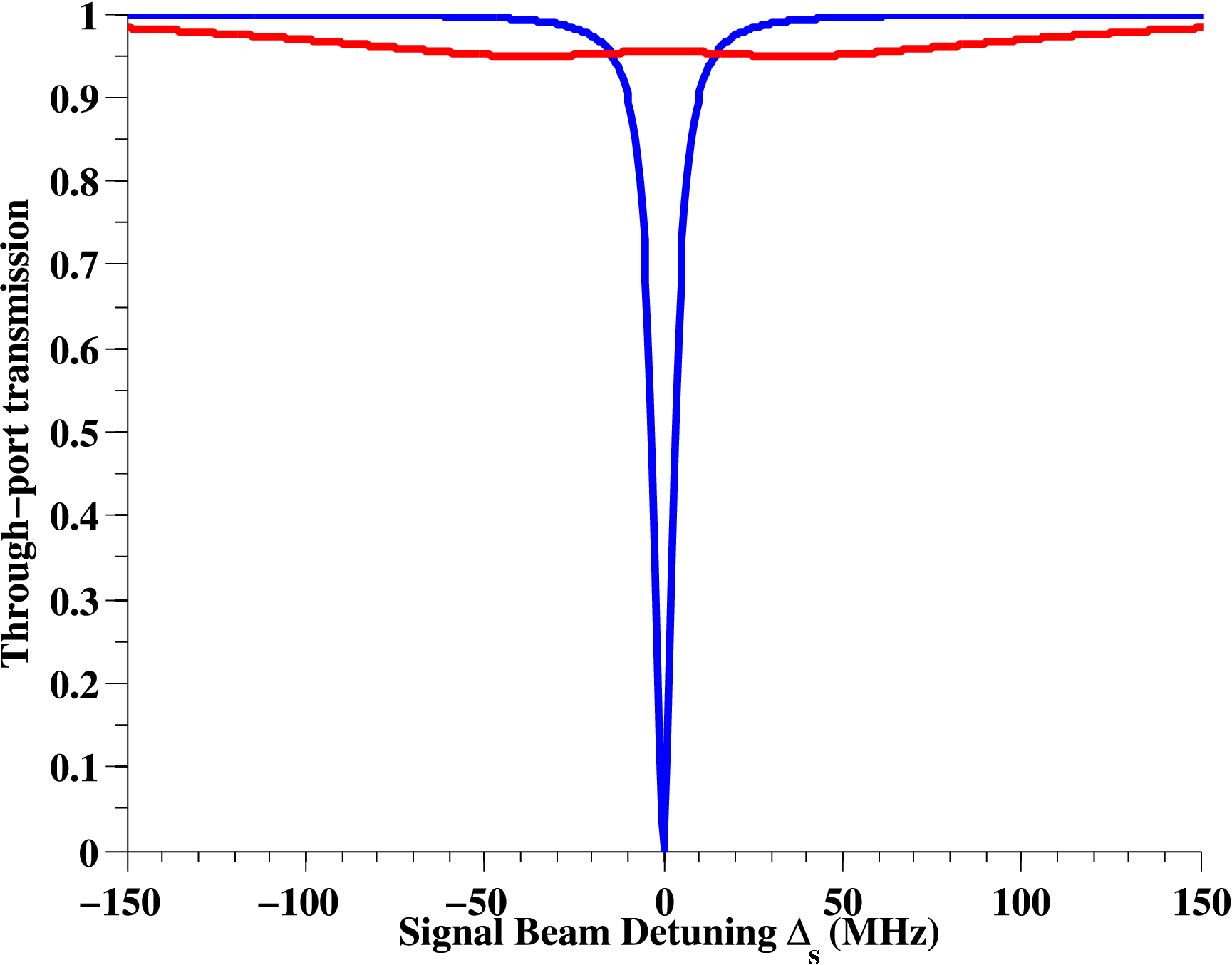}
\end{overpic}
\label{fig:LowerControlWeak-ThroughPort}
}
\subfigure[Drop Port: $P_1/P_2 = 0.1$]{
\begin{overpic}[width=3.99cm]{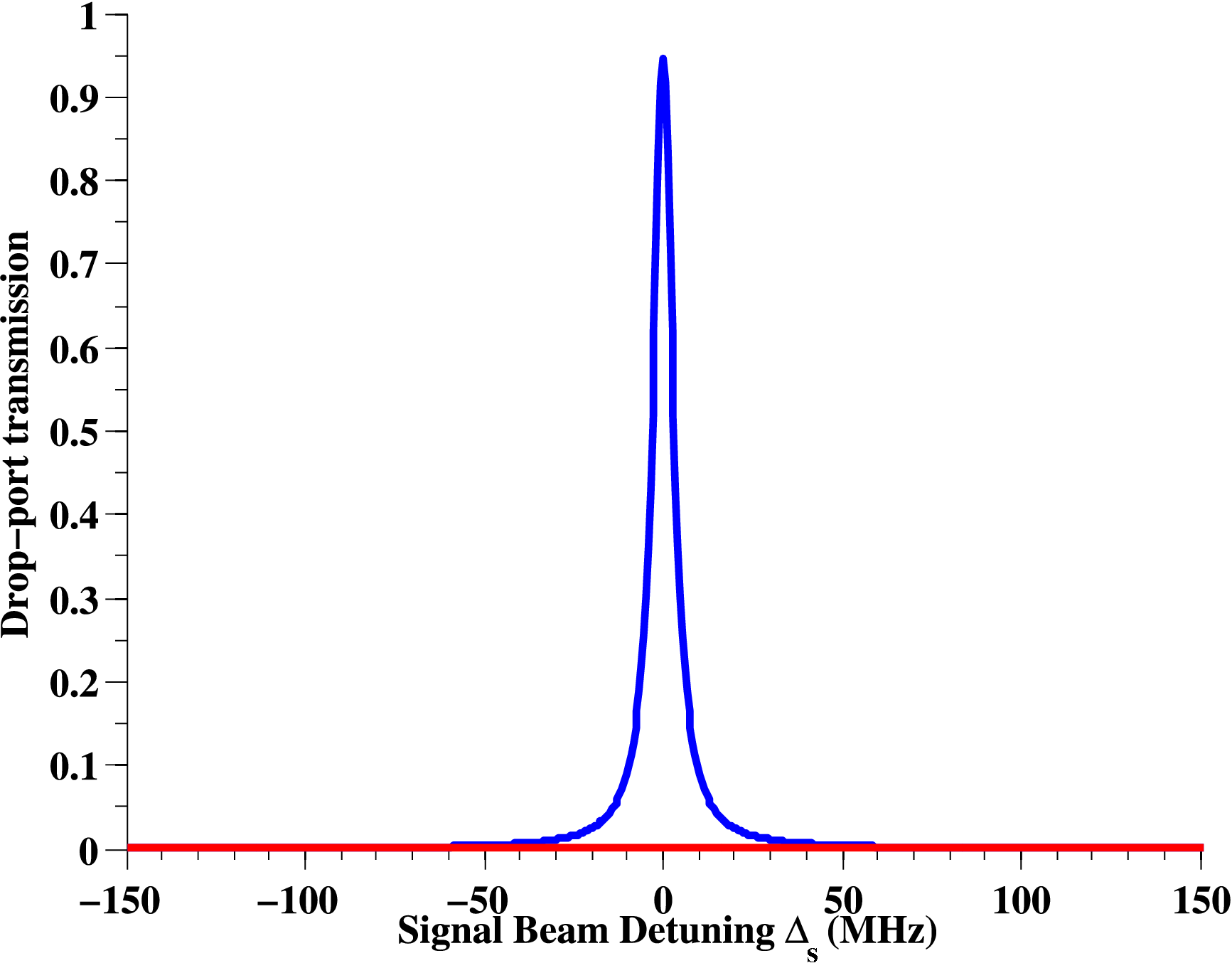}
\end{overpic}
\label{fig:LowerControlWeak-DropPort}
}
\end{center}
\caption{\label{fig:LowerControl}Transmission plots of the through-port (left column) and drop-port (right column) for equal power inputs (top row) and unequal power inputs (bottom row). The transmission plots are for the $\Omega_2$ beam, in which we assume the $\Omega_1$ beam enters the resonator first and is allowed to build up to steady state.  In the bottom row, we demonstrate that a low power input can switch a stronger power input.\vspace{10pt}}
\end{figure}

\begin{figure}[h!]
\begin{center}
\subfigure[Through Port: $P_1/P_2 = 1$]{
\begin{overpic}[width=3.99cm]{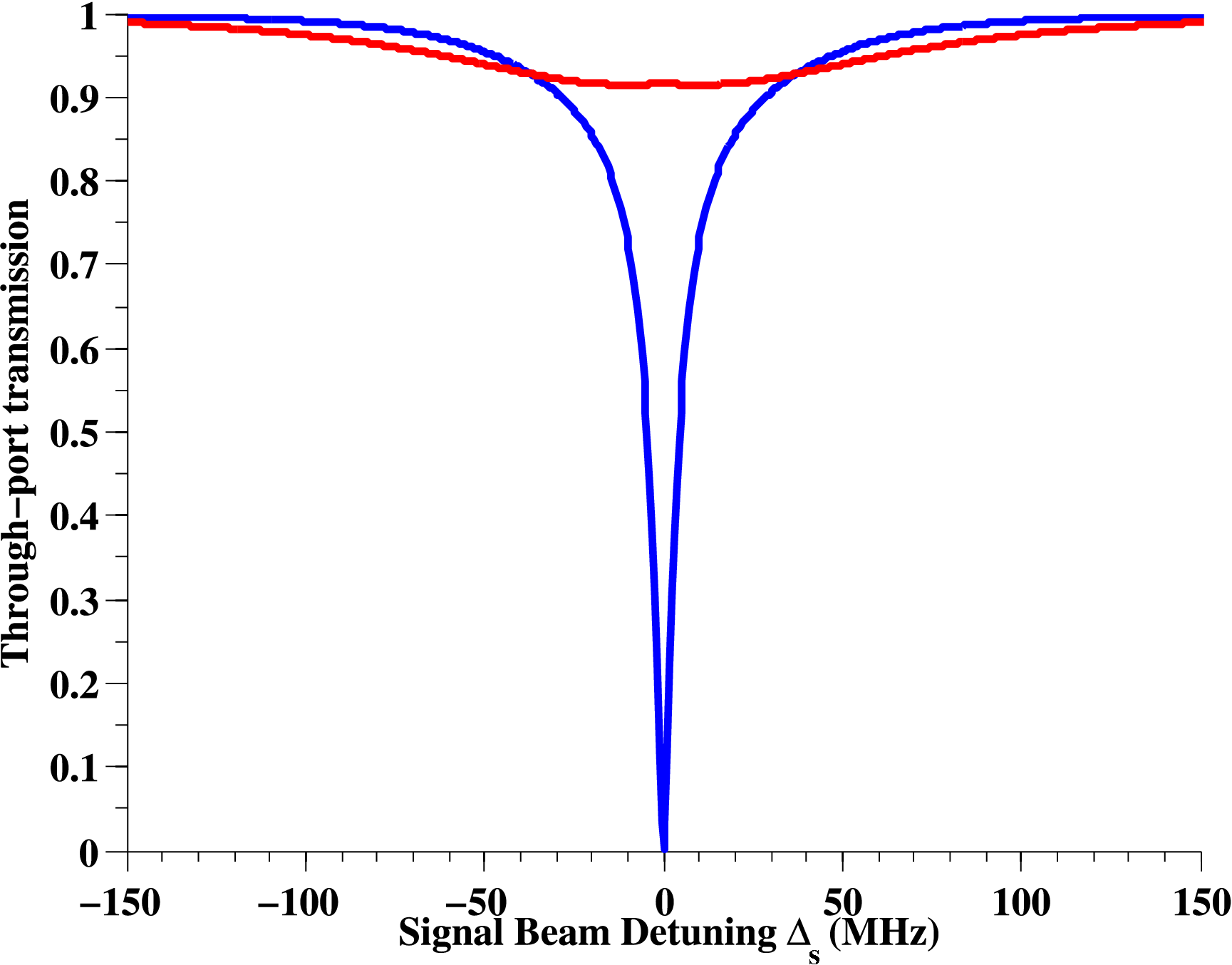}
\put(40,85){{\bf 780 nm Control - 795 nm Signal}}
\end{overpic}
\label{fig:UpperControlEqual-ThroughPort}
}
\subfigure[Drop Port: $P_1/P_2 = 1$]{
\begin{overpic}[width=3.99cm]{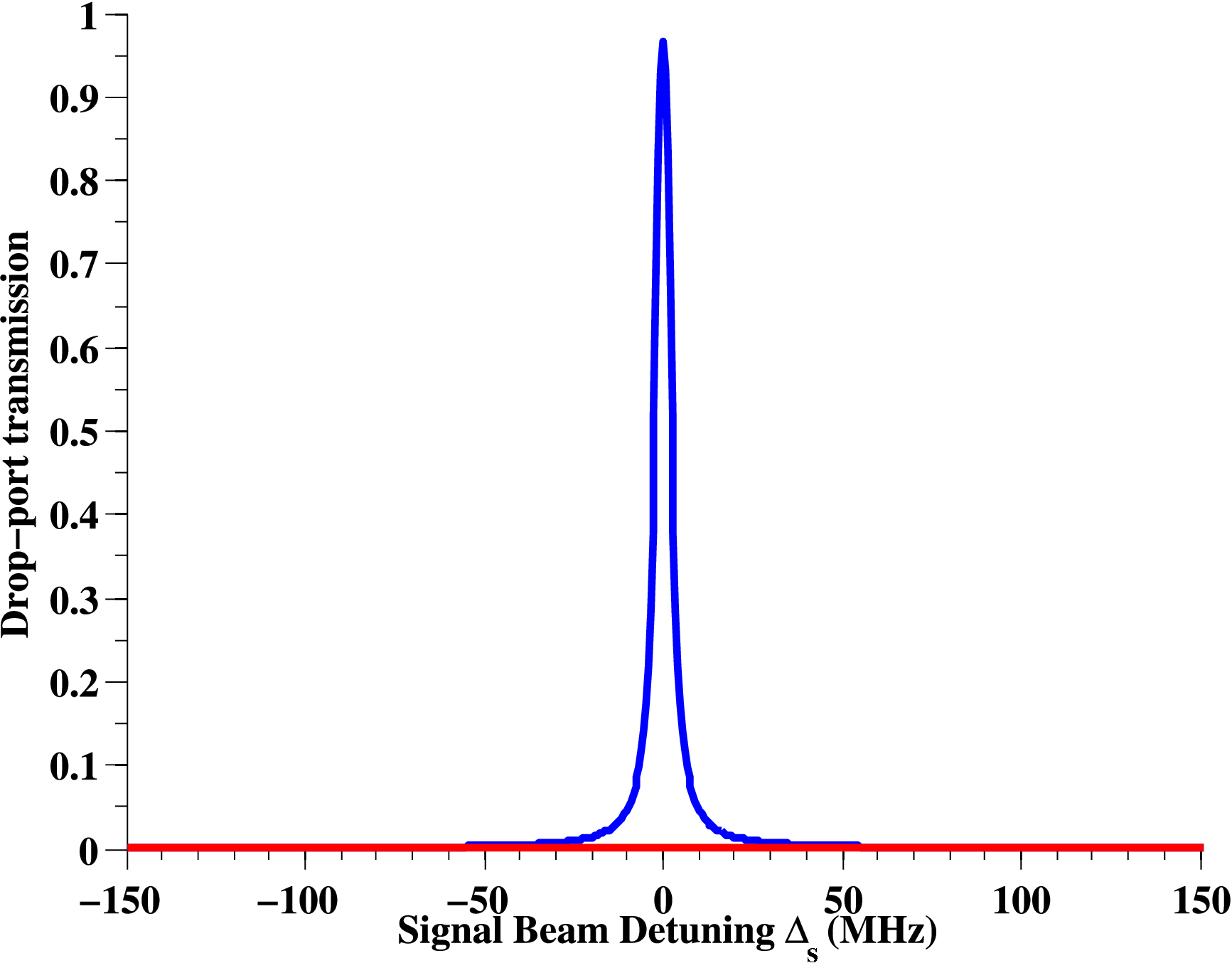}
\end{overpic}
\label{fig:UpperControlEqual-DropPort}
}
\subfigure[Through Port: $P_1/P_2 = 10$]{
\begin{overpic}[width=3.99cm]{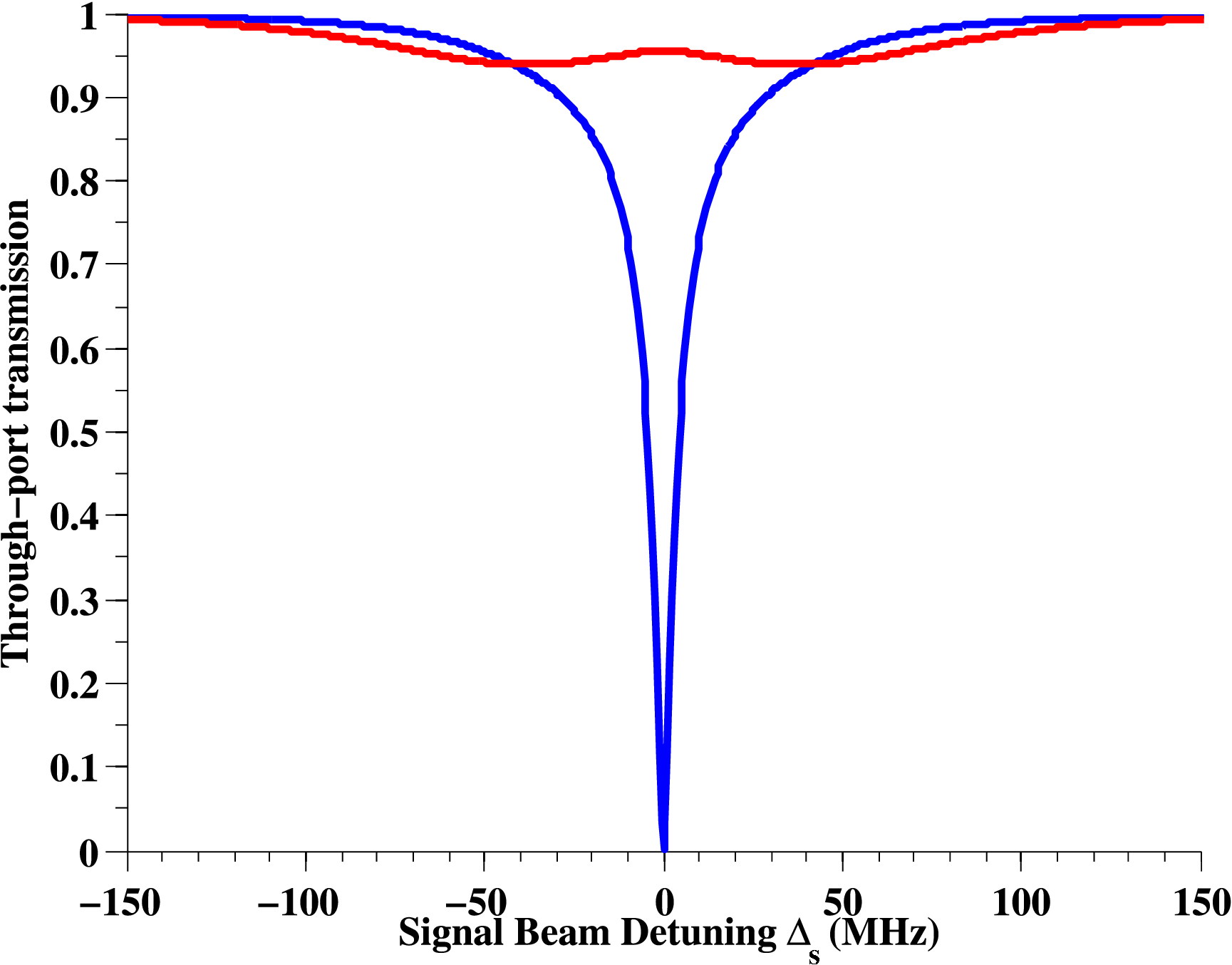}
\end{overpic}
\label{fig:UpperControlWeak-ThroughPort}
}
\subfigure[Drop Port: $P_1/P_2 = 10$]{
\begin{overpic}[width=3.99cm]{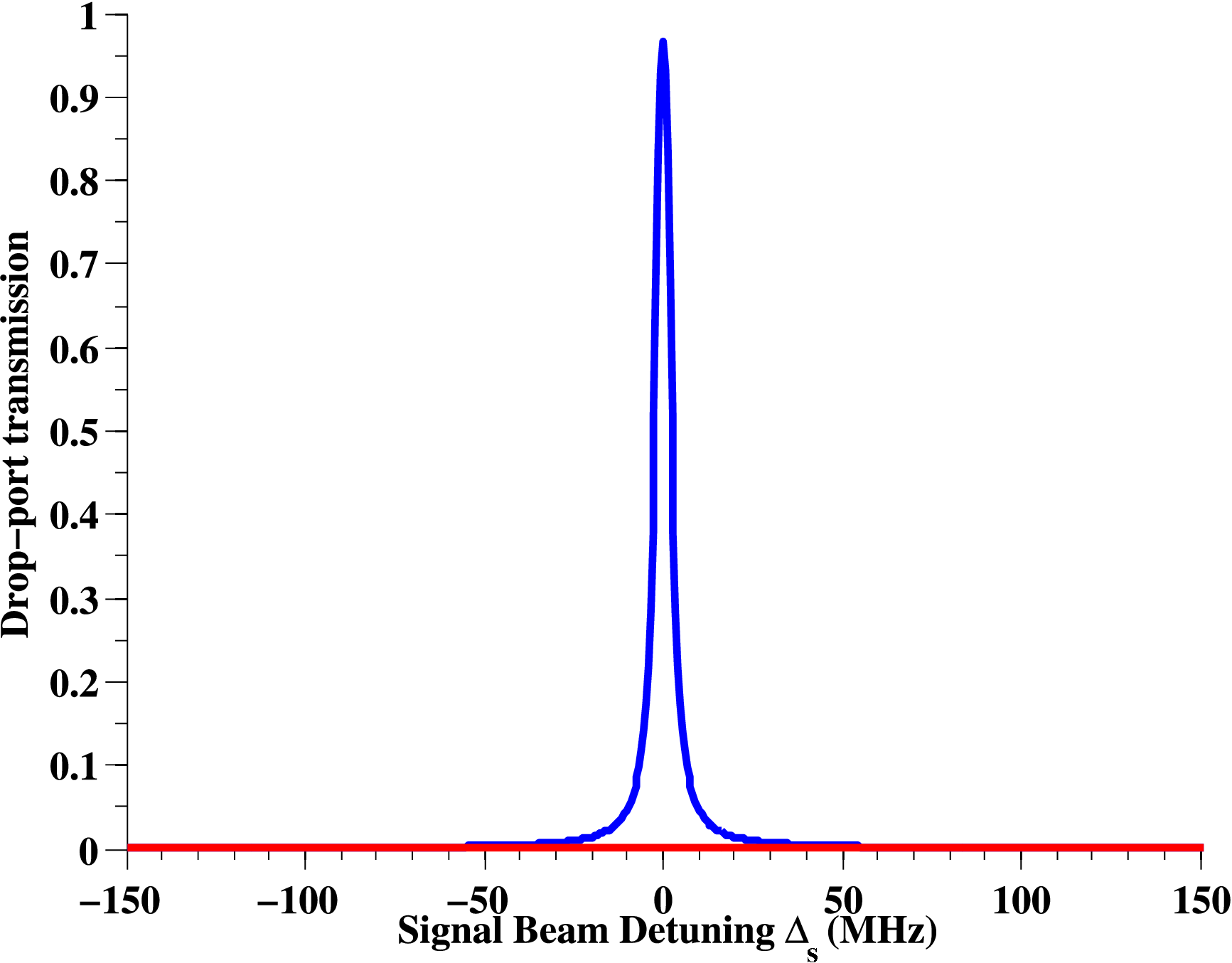}
\end{overpic}
\label{fig:UpperControlWeak-DropPort}
}
\end{center}
\caption{\label{fig:UpperControl}Transmission plots of the through-port (left column) and drop-port (right column) for equal power inputs (top row) and unequal power inputs (bottom row). In contrast to Fig. \ref{fig:LowerControl} these transmission plots are for the $\Omega_1$ beam, in which we assume the $\Omega_2$ beam enters the resonator first and is allowed to build up to steady state.  In the bottom row, we demonstrate that a low power input can switch a stronger power input.}
\end{figure}

\begin{table}[h!]
  \centering
  \caption{\label{tab:results_795}Performance Results - 795 nm Control}\begin{tabular}{r|cc}\hline
   		 & Equal Power & Weak Control \\ \hline
    Drop Port Contrast & 39 dB & 32 dB  \\
    Through Port Contrast & 31 dB & 31 dB \\
    Drop Port Loss & 0.4 dB & 0.4 dB  \\
    Through Port Loss & 0.1 dB & 0.2 dB \\ \hline
  \end{tabular}
\end{table}

\begin{table}[h!]
  \centering
  \caption{\label{tab:results_780}Performance Results - 780 nm Control}\begin{tabular}{r|cc}\hline
   		 & Equal Power & Weak Control \\ \hline
    Drop Port Contrast & 27 dB & 33 dB  \\
    Through Port Contrast & 34 dB & 34 dB \\
    Drop Port Loss & 0.3 dB & 0.3 dB  \\
    Through Port Loss & 0.4 dB & 0.2 dB \\ \hline
  \end{tabular}
\end{table}

The interaction of the evanescent fields with the Rubidium surrounding the cavity is the source of the added loss $\kappa_e$ in the waveguide--resonator coupling model.  To give performance estimates for the switch, we calculate the average absorption coefficient of the signal beam in the resonator from a steady-state solution to the atomic density matrix equations, as described in Ref. \cite{clader2012}.  We account for Doppler broadening as well as the transverse field profile of the two signal beams in the resonator that we calculate using a numerical mode solver.  We take the \ac{EIT} beam (denoted in green in Fig. \ref{fig:NSystem}) to be coming from free-space, with transverse profile much broader than the cavity diameter to eliminate transverse effects. We neglect Doppler broadening of the \ac{EIT} beam, which is valid if one uses a broadband source, with laser linewidth much broader than the Doppler linewidth such as a diode pumped alkali-vapor laser \cite{Gourevitch:08}.  We take the cavity $Q\sim 10^6$ consistent with experimentally proven designs \cite{barclay:131108}, the atomic density to be $N\sim10^{12}$ cm${}^{-3}$, and the waveguides to be over-coupled to the resonator by a factor of 30.

The results of our simulation are shown in Figs. \ref{fig:LowerControl} and \ref{fig:UpperControl}, where we plot the output spectra of the switched beam in both the through-port and drop-ports of the four-port resonator.  The approximate power of the signal beam is 10 pW, while the always-on \ac{EIT} beam is approximately 10 $\mu$W.  In Fig. \ref{fig:LowerControl} we assume that the 795 nm laser, denoted as $\Omega_1$ in the text, is the control beam, while the 780 nm laser is the switched (or signal) laser beam.  We take the opposite designation of control and signal in Fig. \ref{fig:UpperControl}.  With cavity field buildup, 10 pW input power builds to an intensity of roughly 4 W/cm${}^2$ in the cavity which corresponds to only $\sim0.01$ photons on average. 

In all subfigures, we plot the output spectrum of the signal beam where the blue curves indicate the control beam is off, and the red curves are for the control beam turned on.  One can clearly see high-contrast switching occurring near the cavity resonance, for both situations.  The upper row is for the case where both the control and signal are of equal input power, while the lower row is for the situation where the control beam has $1/10$ the power of the signal beam. Performance characteristics for the switch are given in Tables \ref{tab:results_795} and \ref{tab:results_780} for the 795 nm beam as control and 780 nm beam as control cases respectively.  The switching speed is determined by the time it takes for the resonator dynamics to reach steady-state, which imply about 100 ps for the cavity loaded Q's we assume. 

We have theoretically demonstrated a high-speed low-loss all-optical transistor.  We have shown that this scheme should allow for both interchangeability of signal and control, as well as the ability for a weak control beam to switch a strong signal beam, required characteristics for an all-optical transistor. 

We wish to thank Ryan Camacho for the field profile calculation. Funding for this research was provided by IRAD support.


\begin{thebibliography}{10}
\newcommand{\enquote}[1]{``#1''}

\bibitem{1250885}
N.~Kim, T.~Austin, D.~Baauw, T.~Mudge, K.~Flautner, J.~Hu, M.~Irwin,
  M.~Kandemir, and V.~Narayanan, Computer \textbf{36}, 68  (2003).

\bibitem{miller2010optical}
D.~Miller, Nature Photonics \textbf{4}, 3 (2010).

\bibitem{Dawes29042005}
A.~M.~C. Dawes, L.~Illing, S.~M. Clark, and D.~J. Gauthier, Science
  \textbf{308}, 672 (2005).

\bibitem{hu2008picosecond}
X.~Hu, P.~Jiang, C.~Ding, H.~Yang, and Q.~Gong, Nature Photonics \textbf{2},
  185 (2008).

\bibitem{albert2011cavity}
M.~Albert, A.~Dantan, and M.~Drewsen, Nature Photonics \textbf{5}, 633 (2011).

\bibitem{barclay:131108}
P.~E. Barclay, K.~Srinivasan, O.~Painter, B.~Lev, and H.~Mabuchi, Applied
  Physics Letters \textbf{89}, 131108 (2006).

\bibitem{ShahHosseini:10}
E.~S. Hosseini, S.~Yegnanarayanan, A.~H. Atabaki, M.~Soltani, and A.~Adibi,
  Opt. Express \textbf{18}, 2127 (2010).

\bibitem{PhysRevA.82.031804}
J.~Hofer, A.~Schliesser, and T.~J. Kippenberg, Phys. Rev. A \textbf{82}, 031804
  (2010).

\bibitem{jacobs2009all}
B.~C. Jacobs and J.~D. Franson, Phys. Rev. A \textbf{79}, 063830 (2009).

\bibitem{ZenoSwitch}
S.~M. Hendrickson, C.~N. Weiler, R.~M. Camacho, P.~T. Rakich, A.~I. Young,
  M.~J. Shaw, T.~B. Pittman, J.~D. Franson, and B.~C. Jacobs, arXiv:1206.0930v1
   (2012).

\bibitem{harris:36}
S.~E. Harris, Physics Today \textbf{50}, 36 (1997).

\bibitem{PhysRevLett.102.203902}
M.~Bajcsy, S.~Hofferberth, V.~Balic, T.~Peyronel, M.~Hafezi, A.~S. Zibrov,
  V.~Vuletic, and M.~D. Lukin, Phys. Rev. Lett. \textbf{102}, 203902 (2009).

\bibitem{Zhang:07}
J.~Zhang, G.~Hernandez, and Y.~Zhu, Opt. Lett. \textbf{32}, 1317 (2007).

\bibitem{clader2012}
B.~D. Clader, S.~M. Hendrickson, R.~M. Camacho, and B.~C. Jacobs,
  arXiv:1208.4813 [quant-ph]  (2012).

\bibitem{PhysRevLett.81.3611}
S.~E. Harris and Y.~Yamamoto, Phys. Rev. Lett. \textbf{81}, 3611 (1998).

\bibitem{Yan:01}
M.~Yan, E.~G. Rickey, and Y.~Zhu, Opt. Lett. \textbf{26}, 548 (2001).

\bibitem{Mulchan:00}
N.~Mulchan, D.~G. Ducreay, R.~Pina, M.~Yan, and Y.~Zhu, J. Opt. Soc. Am. B
  \textbf{17}, 820 (2000).

\bibitem{PhysRevA.70.022107}
S.~G. Schirmer and A.~I. Solomon, Phys. Rev. A \textbf{70}, 022107 (2004).

\bibitem{PhysRevA.71.022501}
P.~R. Berman and R.~C. O'Connell, Phys. Rev. A \textbf{71}, 022501 (2005).

\bibitem{haus1984}
H.~A. Haus, \emph{Wave and Fields in Optoelectronics} (Prentice-Hall, Englewood
  Cliffs, NJ, 1984).

\bibitem{Gourevitch:08}
A.~Gourevitch, G.~Venus, V.~Smirnov, D.~A. Hostutler, and L.~Glebov, Opt. Lett.
  \textbf{33}, 702 (2008).

\end{thebibliography}

\end{document}